# Interface high temperature superconductivity

Lili Wang, Xucun Ma and Qi-Kun Xue

*State Key Laboratory of Low-Dimensional Quantum Physics, Department of Physics, Tsinghua*

*University, Beijing 100084, China*

*Collaborative Innovation Center of Quantum Matter, Beijing 100084, China*

Email: qkxue@mail.tsinghua.edu.cn

Cuprate high temperature superconductors consist of two quasi-two-dimensional (2D) substructures: $CuO_2$ superconducting layers and charge reservoir layers. The superconductivity is realized by charge transfer from the charge reservoir layers into the superconducting layers without chemical dopants and defects being introduced into the latter, similar to modulation-doping in semiconductor superlattices of AlGaAs/GaAs. Inspired by this scheme, we have been searching for high temperature superconductivity in ultrathin films of superconductors epitaxially grown on semiconductor/oxide substrates since 2008. We have observed interface enhanced superconductivity in both conventional and unconventional superconducting films, including single atomic layer films of Pb and In on Si substrates and single unit cell (UC) films of FeSe on $SrTiO_3$ (STO) substrates. The discovery of high temperature superconductivity with a superconducting gap of ~20 meV in 1UC-FeSe/STO has stimulated tremendous interest in superconductivity community, for it opens new avenue for both raising superconducting transition temperature and understanding the pairing mechanism of unconventional high temperature superconductivity. Here, we review mainly the experimental progress on interface enhanced superconductivity in the three systems mentioned above with emphasis on 1UC-FeSe/STO, studied by scanning tunneling microscopy/spectroscopy, angle-resolved photoemission spectroscopy and transport experiments. We discuss the roles of interfaces and possible pairing mechanism inferred from those studies.

## 1. Introduction

Superconductivity, as a remarkable macroscopic quantum phenomenon, was discovered a centrury ago [1]. It is characterized by disappearance of electric resistance and complete expulsion of magnetic field below a critical temperature $T_C$, and has been observed in many materials. More than five decades ago, Bardeen, Cooper and Schrieffer constructed the microscopic theory of superconductivity, known as BCS theory [2]. According to the BCS theory, electrons in a superconductor collectively bind into "Cooper" pairs and simultaneously condense in much the same way as bosons condense into a superfluid state. The binding interaction is coupling between electrons and vibration of the lattice (phonons). The scheme normally leads to an isotropic s-wave pairing of electrons with opposite momenta near the Fermi surface ($E_F$). The pairing strength, which is characterized by an energy gap $\Delta$, determines $T_C$ with the relation of $2\Delta \sim 3.53 k_B T_C$, where $k_B$ is the Boltzmann constant. With quantitative experimental confirmation of electron-phonon (*e-ph*) coupling in elementary superconductors in mid-1960s [3, 4], the BCS theory has been accepted as a standard theory for conventional superconductors.

According to the modified BCS theory [2],

$$T_C = 1.14 \Theta_D \exp[-1/N(E_F)V], \tag{1}$$

where $\Theta_D$ is the Debye temperature, $N(E_F)$ the electron density of states at $E_F$, and $V$ the attractive electron-electron interaction, in principle, $T_C$ can be quite high by increasing one or more of the three parameters. Unfortunately, not all of the three parameters are completely independent. Ceramic materials have a large $\Theta_D$ but very small $N(E_F)$, thus insulating unless heavily doped. Most elementary metals superconduct but the $T_C$ is low because of small $\Theta_D$. $T_C$ had ever reached 23 K in $Nb_3Ge$ in 1973 [5], and then to 39 K in $MgB_2$ in 2001 [6], which is probably the upper limit of $T_C$ for a typical conventional superconductor [7]. It had been generally accepted by the community that there is no much room for further raising $T_C$.

A new era of superconductivity research was ushered in with the discovery of $T_C$ well above 30 K in La-based cuprate in 1986 [8] and its rapid raising to a temperature well above the boiling point (77 K) of liquid nitrogen in Y-based and Hg-based cuprates in the early 1990s [9, 10]. Historically, cuprates have been coined as unconventional superconductors since their high $T_C$ cannot be explained directly by the BCS theory for conventional superconductors although it is not fully proved. The second upsurge in unconventional superconductivity is the discovery of iron-based superconductors in 2008; $T_C$ of 26 K was first reported in $LaFeAsO_{1-x}F_x$ [11] and the record of 55 K for bulk iron-based superconductor in SmO$_{1-x}$F$_x$FeAs in the same year [12]. The term "unconventional" is notably justified by the primary $d_{x^2-y^2}$ pairing symmetry (with nodes in superconducting gap) [13, 14] and pseudogap state [15-17] for cuprates, possible $s_\pm$ pairing symmetry for iron-based superconductors [18, 19], and a BCS ratio ($2\Delta/k_B T_C$) much larger than 3.53, which goes beyond the conventional wisdoms in the framework of BCS theory. Although the contribution of *e-ph* coupling to high temperature superconductivity has been revealed by various experimental studies such as angle-resolved photoemission spectroscopy (ARPES) [20], inelastic scattering [21] and tunneling spectroscopy [22], the dominant view of the community is that the *e-ph* interaction is, at most, peripherally relevant to high temperature superconductivity. Most researchers believe that the high temperature superconductivity in cuprates and iron-based pnictides and chalcogenides originates from strong electron-electron correlation and that the pairing is mediated by spin fluctuation or short-range magnetic exchange interaction [14-16, 18]. However, the exact picture remains elusive and one of the foremost open problems in condensed matter physics.

In spite of diverse microscopic electronic structures and phenomena, all the compounds with $T_C$ higher than ~30 K, including both unconventional superconductors and conventional BCS superconductors such as MgB$_2$, have a layered structure. Cuprates and iron-based superconductors consist of two types of quai-2D substructures: the supercondcuting layer (CuO$_2$-layer or FeAs/FeSe-layer) and charge reservoir layers, for example, La(Sr)O layer and LaO(F) layer as schematically shown in Fig.1 (a) and (b), respectively [14, 23]. The superconductivity occurs within the superconducting layer with its charge transferred from the reservoir layers, similar to the modulation-doping in semiconductor heterostructures [24]. In MgB$_2$, the characteristic graphite-like 2D boron layers sandwich the triangular Mg layers forming a structure similar to the intercalated graphite [25], and the supercondcutivity can be viewed as having resulted from charge transfer from Mg to graphite-like boron layers as well.

Learning from LaFeAsO$_{1-x}$F$_x$ with $T_C$ = 26 K [11], we realized that the interface between the superconducting layer and the charge reservoir layer may play crucial role in high temperature superconductivity. In mid-2008, we took the plunge and commenced our study on interface superconductivity. The theoretical prototype of interface superconductivity can be traced to "surface superconductivity" proposed by Ginzburg in 1964 [26]. However, the experimental research in this direction just got into stride with the development of advanced thin-film depostition techniques that allow atomic layer precision in the last two decades. We recommend the excellent review by Pereiro *et al.* [27]

and Gariglio *et al.* [28] on the superconductivity of various hetrostructures including PbTe/PbS, LaAlO$_3$/SrTiO$_3$, La$_{2-x}$Sr$_x$CuO$_4$/La$_2$CuO$_4$ and *etc*. We began with searching for high $T_C$ in the heterostructure of ultra-thin metal/oxide films epitaxially grown on semiconductor/oxide substrates. We anticipated to achieve significantly enhanced $T_C$ by taking advantage of high $N(E_F)$ of metal or 2D carrier gas/liquid formed at the interface, high Debye temperature of semiconductor/oxide, and probably strong electron-phonon coupling at interface. That is, we simultaneously maximize the three parameters in Formula (1) by ultilizing interface effect (Fig. 1(c)).

We prepared single atomic layer films of superconductors on semiconductor/oxide substrates by state-of-the-art molecular beam epitaxy (MBE) and studied their superconductivity by combined scanning tunneling microscopy/spectroscopy (STM/STS) and transport measurements. We observed superconductivity in single atomic layer films of both conventional superconductors (Pb and In on Si [29] and Ga on GaN [30]) and high temperature superconductor in 1UC FeSe films on SrTiO$_3$ (STO) (refereed as 1UC-FeSe/STO hereafter) [31]. In this short review, we summarize the interface enhanced superconductivity in Pb/Si, In/Si and FeSe/STO systems, with focus on FeSe/STO system, which exhibits the highest superconducting transition temperature $T_C$ among all the heterostructure systems discovered so far [32-36].

## 2. Interface enhanced superrconductivity in Pb/Si and In/Si systems

Bulk Pb and In are conventional superconductors with $T_c$ ~ 7.2 K and $T_c$ ~ 3.4 K, respectively[1]. When considering size and dimensional effects, the normal trend is that superconductivity is suppressed when the superconducor is thinner than the size of the electron pairs that form the superconducting state. For example, $T_c$ of ultrathin crystalline Pb films is continuously reduced with decreasing film thikness from 9 to 3 atomic layers [37]. In two dimensional limit of single atomic layer, where electrons are mobile only in the planar direction, thermal and/or quantum fluctuations may disturb the coherent motion of the electron pairs and break the superconductivity. Thus, whether superconductivity exists in single atomic layer had been questioned.

It turns out that superconducting gap appears on both single atomic layer Pb and In on Si(111) substrates [29]. Figures 2(a) and (b) display the schematic structure and atomically resolved STM topograph of the striped incommensurate (SIC) phase Pb on Si(111), respectively. Figure 2(c) shows the tunneling spectra as a function of temperature. Well defined U-gap with zero conductance region and two symmetric sharp coherence peaks centered at $E_F$ can be seen at temperatures below 1.82 K. The temperature evolution of superconducting gap is shown in Fig. 2(d). Fitting the data using the BCS gap function yields Δ=0.35 meV, $T_c$ = 1.83 K and BCS ratio $2\Delta/k_BT_C$ =4.4. Despite the $T_C$ is much lower than the bulk value of 7.2 K [1], the BCS ratio is very close to the value of 4.3 in bulk Pb, suggesting that SIC-Pb/Si is a strongly coupled BCS superconductor.

Single atomic layer In grown on Si (111) with $\sqrt{7} \times \sqrt{3}$ reconstruction is also superconducting, as shown in Fig. 2(e)-(h). The values of Δ, $T_C$ and $2\Delta/k_BT_C$ extracted from the BCS fit are 0.57 meV, 3.18 K and 4.16, respectively. Unlike SIC-Pb/Si, where $T_C$ is strongly suppressed compared with the bulk value, the $\sqrt{7} \times \sqrt{3} - In$ phase has a surprising high $T_C$ close to the bulk value of 3.4 K. More significantly, $\sqrt{7} \times \sqrt{3} -$In/Si has an enhanced BCS ratio of 4.16 in comparision to the bulk value of 3.6, implying that it is transformed into a strongly coupled sueprconductor as SIC-Pb/Si.

The superconductiviy observed above can be interpreted as having resulted from interface effect demonstrated in Fig. 1(c): the carriers in the metal overlayer carry the supercurrent and the *e-ph* interactions that glue the electrons to form pairs are provided by both the metallic layer and the interface. This interpretation is supported by ARPES studies, which demonstrate well-developed 2D free-electron-like metallic bands [29] and enhanced *e-ph* coupling constants (1.07 for SIC-Pb/Si [29] and ~ 1 for $\sqrt{7} \times \sqrt{3} -$In/Si [38]) compared with the corresponding films of several atomic layers thick. Moreover, it was found that the superconductivity is indepenedent of dopant type and doping level of the Si substrates, suggesting that only the metal overlayer and the metal-Si interface are responsible for the occurrence of superconductivity [29]. Later on, two independent *in-situ* transport studies indicate that single atomic layer In (Pb) indeed superconducts at 2.8 K (1.1K) [39, 40]. Meanwhile, our interpretation on the correlation of superconductivity with interfacical *e-ph* coupling is also supported by first principles calculation [41]. The 1UC FeSe films on Nb-STO and on insulating STO substrates exhibit simialr superconducting property, which will be discussed in next section. Our idea of interface enhanced superconductivity as demonstrated in Fig. 1(c), therefore, is verified elementarily.

3. **Interface enhanced superrconductivity in FeSe/STO system**

We then moved to investigate ultrathin films of unconventional superconductors to achieve higher $T_C$ in 2010. We chose FeSe because among all of unconventional superconductors it is structurally simplest and the MBE growth recipie of single crystalline stochiometric compounds had been well established in our group [42, 43]. In this section, we discuss the interface enhanced superconductivity in FeSe/STO system and conclude that interface charge transfer and *e-ph* coupling are essential to the high $T_C$ superconductivity therein. In 3.1, we introduce the growth and structure of FeSe/STO(001). In 3.2, we demonstrate its superconducting property, including superconducting gap Δ and superconducting transition temperature $T_C$. We discuss the interface effects, such as strain, charge transfer and *e-ph* coupling, and their contribution to superconductivity in 3.3. We show interface enhanced superconductivity in other related FeSe/STO systems in 3.4.

3.1 Materials and structure

The bulk β-phase FeSe is a superconductor with $T_c$ ~ 8 K at ambient-pressure [44]. One UC β-phase FeSe consists of two Se layers sandwiching a Fe layer with an in-plane lattice constant of 3.78 Å and an out-of-plane lattice constant of 5.50 Å (Fig. 3(a)). Epitaxial FeSe films with Se-terminated (001) surface can be obtained by co-evaporting Fe and Se under Se rich condition (typical flux ratio of ~1:10) on both graphene [43] and STO [31] substrates. When FeSe is grown on graphene, it forms nearly free-standing islands due to very weak interaction and hence intrinsic properties of bulk FeSe are observed, including the in-plane lattice constant of 3.8 Å, V-shaped gap of ~ 2.2 meV and $T_c$ of ~ 8 K [43, 45, 46]. For one UC FeSe films epitaxially grown on STO(001) substrates, their in-plane structure follows that of STO(001) with a lattice constant of ~3.90 Å [31, 47], 3% expanded compared to bulk FeSe [44]. Correspondingly, it is compressed along c-axis with a reduced Se height $h_{Se}$ (above the Fe layer) of 1.31 Å, ~9% smaller than the bulk value of 1.45 Å, as resolved by atomically resolved scanning transmission electron microscopy (STEM) [48]. The Se-Fe-Se bond angle α is enlarged to 111.4°±0.9°, very close to that of a tetrahedron (~109.47°).

Macroscopic uniform single UC FeSe films can be obtained by MBE, which allows transport measurement down to single UC precision. As shown in Fig. 3(b), the morphology of single UC FeSe films perfectly follows the step-terrace structure of STO(001) substrates. Under the Se rich growth condition, excess Se atoms form Se dimers (Fig. 3(c)) and the as-grown Se-rich FeSe$_{1+x}$ films are not superconducting at 4.2 K [49, 50]. Superconductivity occurs only when the density of Se dimer reduces to approximately 2 per 10×10 nm$^2$ and lower after extensive annealing (Fig. 3(d)). Se vacancies appear with elongated annealing (Fig. 3(d)), but they don't destroy superconductivity [49]. It is worthy noting that neither the lattice structure nor the superconductivity is dependent on the bulk property of STO substrates [31, 49]. Below we mainly show STS and ARPES results on Nb-STO(001) substrates and transport on insulating STO(001) unless otherwise noted.

3.2 Superconducting property: U-shaped gap and enhanced $T_C$

It is striking that the single UC FeSe films on Nb-STO(001) exhibit an overall U-shaped gap of 20 meV (Fig. 3(e)), which is nearly one order of magnitude higher than that of bulk FeSe [31]. In single UC FeSe films on insulating STO(001), we observed similar superconduting-like gap $\Delta$ ~15.4 meV, roughly agreeing with the gap magnitude observed on 1UC-FeSe/Nb-STO(001) within experimental uncertainties, which is still visible at 50.1 K (Fig. 3(g)) [49]. A gap closing temperature $T_{gap}$ ~ 66.8 K is deduced from the linear dependence of zero bias conducatce on temperature (Fig. 3(f)). The above observation suggests that 1UC-FeSe/STO(001) is high temperature superconductor with $T_C$ exceeding the known record $T_C$ ~55 K of bulk iron-based superconductors [12]. Furthermore, in terms of the gap observed on FeSe films on graphene, which is ~ 2.2 meV in magnitude and closes at ~ 8 K [43], and assuming the same BCS ratio, we can even anticipate a transition temperature above the boiling point (77 K) of liquid nitrogen.

The above discovery is surprising, not simply because it suggests high temperatue superconductivity with $T_C$ above 50 K but also because the U-shaped gap hints a conventional s-wave pairing symmetry, hence has stimulated tremendous interest in superconductivity community. Our finding was confirmed soon by several independent ARPES and STM/STS studies, which consistently demonstrate a nodeless gap with magnitude of 14-19 meV depending on sample quality [51-55]. For example, the Xingjiang Zhou group performed the first ARPES investigation and demonstrated superconducting-like gap of 15-19 meV which persists to 65 K [51, 52], agreeing well with the STS results shown in Fig. 3(e)-(g). More significantly, they further revealed that the superconducting-like gap is nearly isotropic around the whole Fermi surface, which matches well the observed U-shaped gap shown in Fig. 3(e) [31]. The Donglai Feng group investigated quasi-particle interference patterns in 1UC-FeSe/Nb-STO(001) and the response to both magnetic and non-magnetic impurities. Consequently, they deduced plain s-wave pairing symmetry based on their finding that the quasi-particle interference patterns are against line nodes and the superconductivity is suppressed by magnetic impurities but doesn't respond to non-magnetic impurities [55]. This s-wave pairing symmetry is reminiscent of phonon-mediated pairing under the scheme of BCS theory, which will be discussed in section 3.3.

Since superconductivity is a macroscopic quantum phenomenon, tremendous efforts have been made on transport study and $T_C$ ranging from 40 K to above 80 K has been reported by *ex-situ* measurement [32-34, 36]. The transport measurement is challenging because the single UC FeSe films are too thin (0.55 nm) to survive in atmosphere. By using amorphous Si and crystalline FeTe as capping layer, we performed the first *ex-situ* transport study and found similar $T_C$ in amorphous-Si/3UC-FeSe/STO [31, 56] and FeTe/1UC-FeSe/STO [32-34]. The schematic and the result of *ex-situ* transport measurement on 10UC-FeTe/1UC-FeSe/STO are shown in Fig. 4. The resistance starts to decrease at 54.5 K and drop completely to zero at 23.5 K (Fig. 4(c)). The onset transition temperature $T_{onset}$, defined as the point that the normal resistance and the superconducting transition curves cross, is above 40 K (Fig. 4(c)), increased fourfold with respect to bulk FeSe [57]. Meanwhile, two-coil mutual inductance measurement revealed the formation of diamagnetic screening at 21 K (Fig. 4(d)), which is consistent with $T_{zero}$ ~ 23.5 K. In addition, two independent *dc* magnetic susceptibility measurements provide more information. The Paul Chu group measured 10UC-FeTe/nUC-FeSe/STO(001) (n=1, 2, 3, 4) samples and demonstrated a Meissner state below 20 K, a mesoscopic superconducting state up to 45 K and collective excitations up to 100 K with nature yet to be determined [33]. The Jian Wang group observed a drop crossover around 85 K in 10UC-FeTe/1UC-FeSe/Nb-STO(001) [34]. Recently, the Yayu Wang group performed *ex situ* two-coil mutual inductance measurement on one total 5UC thick films (2UC-FeSe/2UC-$Fe_{0.96}Co_{0.04}$Se/1UC-FeSe/Nb-STO(001)) capped by Se and revealed an onset of diamagnetic screening at 65 K [36], coinciding with the gap closing temperature determined by previous ARPES [52] and STS studies [49]. The higher transition temperature observed in the latter two cases could be due to the fact that FeSe films on Nb-STO(001) are more uniform than those on insulating STO(001) [31, 49]. Since the second UC and thicker films do not superconduct even at 4.2 K [31], we believe that all the superconducting behavior observed on 1-5 UC FeSe films originates from the first UC FeSe right above STO(001), while the additional FeSe layers serve as protection layers. Similarly, in the case of 3UC FeSe films capped by amorphous Si, we guess amorphous Si mixes into the underlying 1-2 UC FeSe films, so that only the bottom layer remains as FeSe and the observed $T_C$ equals to that of 1-UC FeSe films capped by single crystalline FeTe. For films thinner than 3UC capped by Si, no superconducting transition is observed, the same as the case of Pb films capped by Si [37].

FeTe films do protect FeSe films from oxidation in atmosphere, however, they simultaneously suppress the superconductivity through partially mixing with FeSe [48, 58] and providing additional decay channel to the gluing bosons [59]. The atomic intermixing can be resolved from the atomically resolved STEM image shown in Fig. 4(b), as the top layer Se atoms look similar to the Te layer in contrast and obviously brighter than the bottom layer Se atoms. Thus, to pin down the $T_C$ of 1UC-FeSe/STO(001), one would need to either perform transport and magnetic measurement *in situ*, without exposing the film to air, or find a capping material that does not reduce $T_C$. Recently, Jinfeng Jia group accomplished *in-situ* transport measurement. Remarkably, by using microscopic four-point contact probe technique [35], they found $T_C$ above 100 K [35], suggesting that 1UC-FeSe/STO(001) is the second superconductor system with $T_C$ above 77 K. This is to be confirmed by *in-situ* magnetic measurement.

3.3 Interface effect

The 1UC-FeSe/STO(001) exhibits distinctive superconductivity. In contrast to weakly interacting interface between FeSe films and graphene [43, 45], FeSe/STO(001) interface plays a significant role in the remarkably enhanced superconductivity. Here we summarize several distinct behaviors of this system:

(1) 1UC FeSe films on STO(001) exhibit U-shaped gap of 20 meV (Fig. 3(e)) and the second UC and thicker films do not supercontuct at 4.2 K [31], whereas, FeSe films only thicker than 2UC has V-shaped gap of no more than 2.2 meV when they are grown on graphene [45].

(2) With increasing film thickness, the $T_C$ (measured by *ex-situ* transport) of multilayer FeSe films on STO(001) gradually degrades and reduces to the bulk value of 8 K at 50UC [56], which is again in sharp contrast with the case on graphene where the $T_C$ (extracted from the linear fitting of the temperature dependent zero bias conductance) increases and saturates to 8 K at 8 UC [45].

To elucidate the role of FeSe/STO interface, such as strain, interface charge transfer and interface *e-ph* coupling, various experimental investigations have been performed. As the superconductivity of bulk FeSe is very sensitive to pressure [60], we first consider strain effect. The finding that 1UC FeSe films on STO(001) undergo in-plane expansion and out-of-plane compression, as described in Section 3.1, seems to be consistent with the empirical rule in bulk iron chalocogenides that the superconductivity is enhanced when the Se height $h_{Se}$ is reduced [61]. However, according to the empirical relation between $T_C$ and anion height for bulk iron based superconductors [61], a $T_C$ of ~ 20 K is expected for the Se height $h_{Se}$ of 1.31 Å, which is much lower than the value for 1UC FeSe films on STO. Motivated by such empirical phenomena, Donglai Feng group used Nb-STO/KTaO$_3$ [62] and Nb-BaTiO$_3$/KTaO$_3$ [63] as substrates to achieve FeSe films subjected to even stronger in-plane expansion. However, their ARPES study demosntated that the gap closing temperaure is enhanced only 5 K with an additional 5.5% lattice expansion [63]. Meanwhile, we observed similar superconducting gap of 17-20 meV in 1UC FeSe films both on TiO$_2$ [64] and on STO(110) [65, 66] substrates (See Section 3.4.3 and 3.4.4 for details, respectively), although they possess distinct in-plane structure, i.e. square lattice with the same lattice constants as bulk FeSe *vs.* rectangular lattice with 5% anisotropy. Strain is, therefore, excluded from the critical factors for the interface high temperature superconductivity in FeSe/STO. Such conclusion is further supported by the observation of identical superconducting gap across a domain boundary in 1UC-FeSe/STO(001) despite locally compressed lattice [55]. Furthermore, the lattice variation would significantly affect the antiferromagnetic superexchange interactions between Fe moments [67]. Hence the minor role of strain refutes the notion that antiferromagnetic-interaction/spin-fluctuation plays the dominant role in mediating Cooper pairing in FeSe/STO system.

On the other hand, it is widely believed that interface charge transfer and interface enhanced *e-ph* coupling are essential to the high temperature superconductivity in 1UC-FeSe/STO, for their contribution have been explicitly identified by various experimental investigations and theoretical calculations. Below, we successively discuss these two effects.

3.3.1 Charge transfer

Significant charge tansfer at the FeSe/STO interface, resembling that from carrier reservoir layer to superconducting layer in high temperaure superconducors (Fig.1(a) and (b)), has been revealed by extensive ARPES [51-54], STS [49, 50] and transport studies [49]. For example, as shown in Fig. 5(a), the Fermi surface of supercondcuting 1UC-FeSe/STO(001) consists of only electron-like pockets centered around the Brillouin zone corners with a band bottom lying 60 meV below the Fermi level, indicative of formation of 2D electron gas (2DEG). Compared with the Fermi surface of bulk FeSe, the lack of hole pockets in the Brillouin zone center of 1UC-FeSe/STO(001) implies that the 1UC FeSe films are heavily electron doped. As estimated from the Fermi surface volume, the doping level in the FeSe layer is ~ 0.12 electrons per Fe atom [52], which could originate from the oxygen vacancies in the STO substrates [53, 68] and induced by band bending at the FeSe/STO interface [64, 69]. In another example, as demonstrated by the Hall measurement shown in the inset of Fig. 5(c), the carriers change from p-type to n-type with decreasing temperature, indicating that the superconductivity is dominated by n-type carriers. Moreover, the superconductivity transition shifts to higher temperature with more electrons injected into FeSe films from STO under electrical field (Fig. 5(c)). It is evident that the interface charge transfer from STO to FeSe films indeed plays a role in promoting the high temperature superconductivity.

On the contrary, with hole pockets appearing in the Brillouin zone center (Fig. 5(b)) and becoming stronger with increasing thickness [53], 2UC and thicker FeSe films on STO(001) prepared by the same method don't exhibit any signature of superconductivity [53, 70]. Comparing with 1UC-FeSe/STO(001) [52], one may speculate that the absence of superconductivity in multilayer FeSe films is due to insufficient carrier transfer from STO substrate. Recently, it turns out that multilayer FeSe films convert to superconducting when they are doped with sufficient electrons. This is first reported by Y. Miyata *et al.*, as their variable temperature ARPES study showed that 3UC FeSe films become electron populous with potassium (K) adsorption and exhibit a pairing formation temperature of 48 ±3 K at optimal doping [71]. Almost during the same period, we systematically investigated the K adsorption on 1-4UC FeSe films by STS, while the Xingjiang Zhou group and the Donglai Feng group performed ARPES investigation on the surface doping effect with K adsorption.

We found that K adsorption does induce superconductivity in the topmost layer of multilayer FeSe films (> 2UC). However, the gap size is thickness dependent [47]. Potassium atoms adsorb randomly on the surface of FeSe films (Fig. 6(a)) and form local 2×2 and √5×√5 reconstructions at the K coverage of ~ 0.2 ML (Fig. 6(b)). As shown in Fig. 6(c) and 6(d), for all the 2-4UC FeSe films, superconducting gap appears at the K coverage of approxiamate 0.1 ML, increases with further adsorption of K atoms, and reaches a maximum value at an optimal K coverage of 0.2-0.25 ML (roughly corresponding to a charge transfer of ~ 0.1-0.12 electron/Fe). Intriguingly, the gap size decreases with increasing thickness, that is, the superconductivity enhancement degrades with the superconducting layer moving away from the interface. For example, at optimal doping, 2UC, 3UC and 4UC films exhibit superconducting gap of 14.5 ±1.0 meV, 13.1 ±1.4 meV and 11.9 ±1.4 meV, respectively (Fig. 6(d)). In contrast, FeSe films thicker than 4UC under optimal doping universally exhibit a little smaller gap of 10 meV persisting up to 44 ± 2 K [72, 73]. Given that K atoms mainly dope electrons into the topmost FeSe layer [72, 74] and the doping level can reach and even exceed the value of ~0.1-0.12 electron/Fe transferred from STO to 1UC FeSe films, the finding that the superconducting gap gradually decays with increasing thickness up to 4UC and remains constant for thicker films suggests that STO substrate contributes additional role in promoting the superconductivity, that is, interfacial *e-ph* coupling that will be discussed in Section 3.3.2.

We further found that the magnitude of the superconducting gap observed on K-coated 2UC FeSe films depends on the superconducting property of the underlying 1UC FeSe films [75]; it reaches a maximum gap of ~ 17 meV and ~ 11 meV depending on the underlying 1UC FeSe is superconducting and not, respectively. The underlying 1UC FeSe films can be superconducting or nonsuperconducting, which can be achieved by controlling the annealing temperature above or below 450°C, respectively [75].

The gap size in the latter case is consistent with the value of 10 meV achieved on K-coated FeSe films thicker than 4UC [72]. If the gaps of 10-11 meV are solely induced by electron doping, the further enhancement to 17 meV when the underlying 1UC films are superconducting could be attributed to the additional role of FeSe/STO interface. For 2UC FeSe films, the additional enhancement factor is ~ 55% (from 11 mev to 17 meV). In the case of 1UC FeSe films, they undergo stronger suppression induced by thermal/quantum fluctuation in the 2D limit, since K-coated 1UC-FeSe films on graphene exhibits exclusively small gap of 6.6 meV in comparison with ~10 meV for K-coated 2-UC and thicker films on the same substrate [73, 76]. If compared with K-coated 1UC-FeSe/graphene, the gap size of 1UC-FeSe/STO(001) [31] is impressively increased threefold.

Very recently, a high $T_C$ of 35-48 K [77-79] and a superconduting gap of 9-14 meV [73, 76] were successively achieved in heavily electron doped FeSe thin films/flakes by using liquid-gating technique and K-coated FeSe films on graphene, respectively. Moreover, this high temperature superconductivity occurs after the native low $T_C$ superconductivity ($T_C$ ~ 8, V-shaped gap of ~2.2 meV) vanishes completely with increasing doping, showing up as a separated phase [76, 78]. The native low $T_C$ superconductivity disappears with an extra electron doping of ~0.02 $e$ per Fe atom, and the high temperature superconductivity occurs and reaches optimal at doping of ~0.04 and ~0.12 $e$ per Fe, respectively [76]. This finding disclosed that the non-superconducting behavior of multilayer FeSe films on STO(001) is due to weaker electron doping from STO (~0.02 $e$ per Fe atom for the second UC films) which suppresses the low $T_C$ superconductivity but is insufficient to boost high $T_C$ superconductivity. On the other hand, 1UC-FeSe/STO(001) is directly promoted to the high $T_C$ regime with a large amout of charge tansfer from STO (0.12e/Fe), which correlates to the double-TiO$_x$ layer (Fig. 4(b)) at FeSe/STO interface [48, 80]. The above results clearly indicate that the charge transfer from STO substrates to FeSe films plays a crucial role.

3.3.2 *e-ph* coupling

Compared with all the other electron-doped or alkali-metal intercalated FeSe superconductors, 1UC-FeSe/STO(001) exhibits following unique features:

(1) The 1UC-FeSe/STO(001) possesses a $T_C$ of ~20 K higher and a gap of at least 5 meV larger than the corresponding values of all the other FeSe correlated superconductors [76, 78, 81-83].

(2) Compared with alkali-metal intercalated FeSe superconductors, for example, (Tl,Rb)$_x$Fe$_{2-y}$Se$_2$ superconductor ($T_C$ = 32 K, effective electron mass ~6.1m$_e$) [84], the electron correlation in 1UC-FeSe/STO(001) is weaker (effective electron mass ~2.7$m_e$) [51]. This fact, together with (1), refutes the notion that the high $T_C$ in 1UC-FeSe/STO(001) is due to a stronger electron-electron interaction.

(3) In contrast to the competition between superconductivity and nematic order observed in multilayer films (> 2UC)[72, 85], neither xz/yz band splitting signaling the nematic order [31, 51, 54] nor any sign of strong nematic fluctuation [55] is observed in 1UC-FeSe/STO(001).

(4) The superconductiviy in 1UC-FeSe/STO(001) is always continuously suppressed with K adsorption (Fig. 6(c) and 6(d)), independent of the initial state (superconducting or not) [75]. On the contrary, the supconductivity can be further enhanced when electrons are injected from STO with field effect (Fig. 5(c)) [49].

The features (1)-(3) imply that the FeSe/STO interface contributes additional profound effect besides doping and hence provides additional enhancement in superconductivity. The additional role could be the interface enhanced *e-ph* coupling, as we proposed [31], which has been experimentally identified by successive APRES [54], STS [47], ultrafast optical spectroscopy [59] and high resolution electron energy loss spectroscopy (HREELS) [86] investigations and theoretically supported [87-91]. First, ARPES observation of "replica" bands with energy seperation of ~ 100 meV in 1UC-FeSe/STO(001) suggests the coupling of FeSe electrons with oxygen optical phonons in STO at such frequency [54]. Furthermore, an *e-ph* coupling constant λ ~ 0.5 is obtained, as estimated from internsity ratio of the replica bands. Second, in agreement with previous ARPES [51, 52, 54] and STS [31, 49] results, $T_C$ = 68(-5/+2) K, Δ = 20.2 ± 1.5 meV and λ = 0.48 were revealed by ultrafast optical spectroscopy study [59]. The *e-ph* coupling constant λ = 0.48 is triple the value of 0.16 measured by the same method for bulk FeSe [92]. Third, a Fuchs-Kliewer (F-K) phonon at frequency of ~ 92 meV was recently revealed by surface phonon investigation using HREELS, consistent with previous ARPES results [54] in terms of phonon energy. More importantly, the electric field generated by this F-K phonon decays exponentially with a decay length of 2.5 UC FeSe [86], which agrees with the observed thickness-dependent superconducting gap shown in Fig. 6(d). These findings point out that the oxygen longitudinal optical (LO) mode of STO with energy reaching ~100 meV (partially) contributes to the additonal enhancement beyond doping. This special oxygen LO phonon is demonstrated to be a special oxygen-vacancy induced flat phonon mode, mainly composed of relative Ti and O atomic displacements along [001] direction in the top two layers of STO substrate [88]. Meanwhile, a quantum Monte Carlo computation showed that this interface *e-ph* coupling can significantly enhance the pairing strength, irrespective of the pairing symmetry as well as its underlying electronic origin [91].

Last but not least, the significant change in the *e-ph* interaction in FeSe films owing to the coupling with STO substrates is also identified in tunneling spectra; the signature of *e-ph* coupling, *i.e.* symmetric dip-hump features, shows up. As displayed in the normalized tunneling spectra shown in Fig. 7(a) and Fig. 7(b), the emergency of the superconducting gap is accompanied with two pairs of dip-hump features, which degrade simultaneously with increasing temperature [47]. Displayed in Fig. 7(c) is the phonon mode energy $\Omega$, extracted from the second derivative of tunneling conductance, as a function of superrconducting gap. The energy distribution of the phonon modes collapses basically into two distinct groups centered at 11.0 meV ($\Omega_1 = 11.0 \pm 2.1$ meV) and 21.5 meV ($\Omega_2 = 21.5 \pm 4.5$ meV), despite the superconducting gap changing significantly from 6.5 meV to 19 meV (depending on the film thickness and K coverage). On the other hand, for ultra-thin films of FeSe grown on and weakly bonded to graphene, although a bosonic mode was observed, the energy is much smaller and only in a level of 2.7−4 meV [93]. The sharp contrast again proves the special role of the STO substrate in boosting the *e-ph* coupling and hence the high $T_C$ superconductivity.

The identification of the two group phonon modes with energy of ~11.0 meV and ~ 21.5 meV and their contribution to the high $T_C$ superconductivity in 1UC-FeSe/STO(001) is further supported by first-principles calculation [90]. According to the description in Ref. 90, one of the roles of STO substrate is to stabilize the 1UC FeSe films to a nearly square arrangement so as to prevent the films from undergoing a shear-type structure transition as the case in bulk [90]. Indeed, the square structure is evidenced from the STM images shown in Fig. 3(d), and signal of nematic order has never been observed in 1UC-FeSe/STO [31, 51, 54], which is in contrast to bulk FeSe [94]. As a result of this substrate-bound structure, two *e-ph* coupling channels with phonon frequencies of 10 meV and 20 meV [90], in excellent agreement with experimental observation shown in Fig.7(c), are opened. This excellent agreement suggests that the STO substrates indeed act as a template and further enhance the *e-ph* interaction from FeSe phonons. The corresponding *e-ph* coupling constant is calculated to be $\lambda = 1.6$ [90], ten times of the value $\lambda = 0.16$ for bulk FeSe [92].

As discussed in section 3.3, with strong electron-electron correlation being excluded as a key role, we show that the interface charge transfer and *e-ph* coupling cooperatively contribute to the high $T_C$ in 1UC-FeSe/STO(001) systems. We roughly estimated the respective contribution of the interface charge transfer and *e-ph* coupling as the former can enhance the gap to 10-14 meV and the latter takes the responsibility for additional enlargment of 6-10 meV. Our initial proposal [31] as depicted in Fig. 1(c) is, therefore, verified to be an efficient approach for raising $T_C$. The evidence of enhanced *e-ph* coupling discussed above, together with the U-shaped gap and s-wave pairing symmetry discussed in Section 3.2, points out that the pairing mechanism in 1UC-FeSe/STO(001) could be rather conventional, *i.e.* pairing mediated by phonons. This conjecture is supported by a recent calculation based on *e-ph* coupling mechanism, which revealed that a $T_C$ of 77 K is obtainable for 1UC-FeSe/STO(001) when the experimental identified LO phonon energy of 100 meV [54, 86], *e-ph* constant of 0.5 [54, 59] and chemical potential of 60 meV [51, 53, 54] are taken into account and the Coulomb repulsion is neglected due to the huge dielectric constant of STO [89]. Nevertheless, the detailed picture of interface *e-ph* coupling and some puzzles still need to be further explored. For example, exploring the isotope effect of $T_C$ will help pin down the pairing mechanism. The interaction between FeSe and double-$TiO_x$ layer terminated STO, which correlates to the significant charge transfer and *e-ph* coupling at the interface, deserves further study. It is also intriguing to observe continuous suppression of the superconductivity with K adsorption on 1UC-FeSe/STO(001), feature (4) listed in section 3.3.2. Given that the interface *e-ph* coupling is assocaited with the formation of interface electric dipole, due to the relative displacement of the Ti cations and the oxygen anions [48, 88], the intrinsic quantum paraelectric/incipient ferroelectric of STO [95-97], and the electric field generated by F-K phonon mode of STO [86], we can interprete the suppression of superconductivity in 1UC FeSe films with electrons doped from topside as having resulted from the counteraction of the interface electric dipoles, which, thus, weakens the screening effect and Cooper pairing strength in turn.

3.4 Interface enhanced superconductivity in related systems

It is natural to ask whether the interface enhanced superconducitvty scenario works in other systems. To test this idea, we have grown 1UC $FeTe_{1-x}Se_x$ films on STO(001), 1UC $K_xFe_2Se_2$ films on STO(001), 1UC FeSe films on STO(110) and on $TiO_2$(001) by MBE and studied the superconducting properties by *in situ* STS and *ex situ* transport. We found that all the systems exhibit remarkably enhanced superconductivity compared with the corresponding bulk materials. Below we briefly discuss the main results in the four systems.

3.4.1 $FeTe_{1-x}Se_x$ films on STO(001)

Similar to FeSe, $FeTe_{1-x}Se_x$ films are prepared by co-depositing Fe, Se and Te and form ordered Se/Te–terminated (001) surface on STO(001) substrate. Displayed in Fig. 8(a) is a typical morphology of 1-UC

FeTe$_{1-x}$Se$_x$ films, where bigger Te atoms are imaged brighter while smaller Se atoms darker. U-shaped gaps with vanishing conductance centered at $E_F$ are observed in all the 1UC FeTe$_{1-x}$Se$_x$ films (x=0.1, 0.3, 0.5, 0.6) and the gap size varies from 12 meV to 16.5 meV depending on the Se composition (Fig. 8(b)). Compared to the superconducting gap ~1.7 meV of the optimally doped bulk FeTe$_{0.6}$Se$_{0.4}$ single crystal [98], the gap size of 1UC-FeTe$_{1-x}$Se$_x$/STO(001) is enlarged at least six times, indicative of interface enhanced superconductivity. In addition, several groups of phonons with frequency of ~10 meV, ~20 meV and ~25 meV, which are consistent with $E_g$(Te/Se), $A_{1g}$(Te/Se)/TO$_2$(STO) and $B_{1g}$(Fe) modes, respectively, are observed. However, the spin resonance mode at ~6 meV as observed in bulk FeTe$_{1-x}$Se$_x$ [99, 100] is not observed on this single UC films. The above results demonstrate the deterministic role of FeTe$_{1-x}$Se$_x$/STO interface in the enhanced superconductivity again. It is worthy noting that in contrast to the filamentary superconductivity reported in bulk FeTe$_{1-x}$Se$_x$ when the Se ratio is smaller than 0.29, the superconducting gap is persistent on the whole surface of 1UC FeTe$_{0.9}$Se$_{0.1}$ films with quite weak position dependence (about 14 meV on Se sites and 12 meV on Te sites), indicating that the interface enhance effect is such strong that supasses the contribution of Te substitution. The resemblance with 1UC-FeSe/STO(001), especially enlarged U-shaped gap and evident *e-ph* coupling, implies that interface engineering is a rather general approach for raising superconductivity temperature. Moreover, 10UC-FeTe/1UC-FeTe$_{0.5}$Se$_{0.5}$/STO(001) and 10UC-FeTe/1UC-FeSe/STO(001) exhibit almost similar R-T behavior and equal $T_{onset}$ ~ 40 K and $T_{zero}$ ~21 K [58], indicative of substitution of Te for Se during the growth of capping layer, consistent with the intermix of Se and Te at the FeTe/FeSe interface as shown in Fig. 4(b).

3.4.2 K$_x$Fe$_2$Se$_2$ films on STO(001)

1UC K$_x$Fe$_2$Se$_2$ films were obtained after the K-coated 2UC FeSe films were appropriately annealed so that the K atoms intercalated between the two FeSe layers [75]. In morphology, 1UC K$_x$Fe$_2$Se$_2$ films are characterized by a step height of 0.7 nm [75] and $\sqrt{2}\times\sqrt{2}$ reconstruction (Fig. 8(c)). As expected, a spatially uniform U-shaped superconducting gap of 14.5 meV is observed (Fig. 8(d)), which is significantly larger than $\Delta$ ~7 meV of bulk K$_x$Fe$_2$Se$_2$ [101], $\Delta$ ~4 meV for K$_x$Fe$_2$Se$_2$ films on graphene [102], and $\Delta$ ~9 meV for thicker K$_x$Fe$_2$Se$_2$ films on STO [103]. Given that bulk K$_x$Fe$_2$Se$_2$ is heavily electron doped, the enhancement observed here should be mainly due to the interface enhanced *e-ph* coupling effect. Meanwhile, the superconducting gap of 14.5 meV in 1UC K$_x$Fe$_2$Se$_2$ films is about 3 meV smaller compared with K-coated 2UC FeSe films at optimal doping, probably owing to the similar counteraction of interface enhance effect with K atoms adsorbed on 1UC FeSe films as shown in Fig. 6(c).

3.4.3 FeSe/STO(110)

Different from STO(001) surface, STO(110) is polar and has anisotropic in-plane lattice- 3.91 Å and 5.53 Å along the [001] and [110] direction, respectively. FeSe films form Se-terminated (001) surface with anisotropic lattice $a = 3.93$ Å and $b = 3.78$ Å along the [100] and [1-10] directions of STO substrate, respectively (Fig. 8(e)). Despite anisotropic lattice structure, the 1UC FeSe films exhibit a gap of 14-17 meV (Fig. 8(f)), similar in magnitude to most reports of superconducting gaps but a little bit smaller than the maximum value (20 meV) in 1UC-FeSe/STO(001) [31, 50]. This result excludes strain as a critical factor and is inconsistent with that antiferromagnetic-interaction/spin-fluctuation plays the dominant role in mediating Cooper paring as we discussed in Section 3.3.

The enhanced superconductivity in 1UC-FeSe/STO(110) can be consistently explained in the interface enhanced *e-ph* scenario that we discussed in Section 3.3. Basically, STO(110) substrates have comparable 2D carrier density (due to oxygen vacancies) [104, 105] and similar O-Ti-O stretching mode with energy at ~ 100 meV [63, 106]. Compared with 1UC-FeSe/STO(001), 1UC-FeSe/STO(110) always exhibits a slightly smaller superconducting gap. We speculate that the dielectric constant and the spatial extension of 2D carriers may play a role. On one hand, for the 2D superconductivity, where the electrons are confined in the plane parallel to the FeSe/STO interface and form 2DEG, in-plane dielectric constant $\varepsilon_\parallel$ reflects the strength of electron screening or Coulomb interaction. And the interface *e-ph* coupling constant is proportional to $\sqrt{\varepsilon_\parallel/\varepsilon_\perp}$, where $\varepsilon_\parallel$ and $\varepsilon_\perp$ are the dielectric constant parallel and perpendicular to the FeSe/STO interface, respectively [87]. The finding that FeSe/STO(110) exhibit smaller superconducting gap agrees with the fact that $\varepsilon_\parallel$ ($\varepsilon_\perp$) of STO(110) is smaller (larger) than that of STO(001) [107]. On the other hand, the spatial extension of 2D carriers along the vertical direction is larger for STO(110) surface than for STO(001) surface [108]. This weaker confinement of 2D carriers in STO(110) should give rise to lower charge transfer to FeSe films, therefore, lower $T_C$ and weaker $T_C$ modulation with electrostatic gating [65]. This indicates that stronger confinement of 2D carriers promotes higher $T_C$, agreeing with the quasi-two-dimensionality in high temperature superconductors introduced in Section 1.

3.4.4 FeSe/TiO$_2$(001)

Learning that FeSe bonds to double-TiO$_x$ layer at the FeSe/STO(001) interface (Fig. 4(b)), we grow anatase $TiO_2$ directly on STO(001) as substrate for FeSe growth. The anatase $TiO_2$ is characterized with distinct O-Ti-O triple layered planes with in-plane lattice constant of 3.78 Å, much closer to that of bulk FeSe in comparison with STO(001). Consequently, the 1UC FeSe films epitaxially grown on this anatase $TiO_2$ possess the same in-plane lattice constant as $TiO_2$, 3.80 ±0.05 Å, slightly larger than the value (3.76 Å) of bulk FeSe (Fig. 8(g)). This result suggests nearly strain-free FeSe films formed on anatase $TiO_2$(001), resembling with those on graphene. Intriguingly, the 1UC-FeSe/$TiO_2$(001) exhibits U-shaped gaps of 21 meV in magnitude (Fig. 8(h)) while 2UC-FeSe/$TiO_2$(001) exhibits no superconductivity signature at all, resembling with those on STO(001) and STO(110). Instead of single vortex observed on 1UC-FeSe/STO(001) (due to dense domain boundaries), periodic vortex lattice forms on 1UC FeSe films on $TiO_2$(001) substrates [64]. The above results demonstrate unambiguously the occurrence of high-$T_C$ superconductivity in 1UC-FeSe/$TiO_2$(001) and hence support our conclusion that strain is not crucial and that antiferromagnetic-interaction/spin-fluctuation doesn't play the dominant role as discussed in Section 3.3. Considering the oxygen LO phonon that couples to FeSe electrons and contributes to the interface enhanced superconductivity in 1UC-FeSe/STO(001) is composed of relative Ti and O atomic displacements (see Section 3.3), we speculate that the case of 1UC-FeSe/$TiO_2$(001) holds the same mechanism. This is verified by a recent ARPES investigation of 1UC-FeSe films on Rutile $TiO_2$(001) substrates which reveals the coexistence of replica bands with an energy separation of 90 meV and superconductivity with a $T_C$ of 63 K [109]. The above finding echoes the essential role of oxygen phonons.

4. **Summary and Perspective**

In this brief review, we have discuss the interface enhanced superconductivity in Pb/Si, In/Si and FeSe/STO systems. For FeSe/STO system, the compelling evidence of enlarged superconducting gap (14-20 meV) and enhanced $T_C$ (> 40 K) demonstrates that interface engineering provides a feasible way for rational design and preparation of high $T_c$ superconductors. As we proposed previously [31], by fabricating sandwiched heterostructure, for example, 2-3 UC FeSe sandwiched between STO [47] or $TiO_2$ [64, 109, 110] on both sides, much higher $T_C$ may be achieved.

We further demonstrate that interface charge transfer and interface enhanced *e-ph* coupling are essential to the enhanced superconductivity. A related calculation reveals that a $T_C$ of ~ 77 K is possible by a combination of three factors: high LO phonon energy, large *e-ph* coupling constant and huge dielectric constant of the STO substrate suppression the Coulomb repulsion [89]. Most recently, a record high $T_C$ of 203 K was reported in H$_2$S under extreme pressure [111], which is a conventional BCS superconductor. Thus, the discovery of high temperature superconductivity in 1UC-FeSe/STO and H$_2$S strongly suggests that high $T_C$ (well beyond the McMillan limit) is achievable under the BCS scenario.

In terms of the resemblance between the FeSe/STO interface and the built-in multi-interfaces in cuprates and iron pnictide superconductors as shown in Fig. 1, we conjecture that both doping resulted charge-transfer [112] (modulation-doping as in semiconductor heterostructure [24]) and *e-ph* coupling contribute the high temperature superconductivity in cuprates and iron-based pnictides and chalcogenides. The essential role of oxygen phonons, which has been evidenced in both FeSe/STO system [54, 109] and cuprate [20, 21] definitely deserves to be revisited. For searching for high $T_C$ under the scheme of doping-charge-transfer and *e-ph* coupling, we can follow the idea depicted in Fig. 1(c), e.g. fabricating heterostructures of metal and material that has high-energy phonon modes (such as diamond, BN, Al$_2$O$_3$, TiO$_2$ and etc.), or heterostructures involving Mott-Hubbard insulators that can be doped effectively with band-bending effect [112]. If ultra-thin or even monolayer Mott-Hubbard insulator film can be prepared on insulating substrate, electric field effect may be employed to tune it superconducting with high $T_C$. We can further fabricate superlattice of such heterostructures to achieve higher $T_C$, as in the case of the multi CuO-layer cuprates [9, 10]. Experiments in this direction are underway in our lab.

**Acknowledgements**

The authors would thank the collaborations and discussions with Qingyan Wang, Wenhao Zhang, Zhi Li, Hao Ding, Chenjia Tang, Chong Liu, Ding Zhang, Huimin Zhang, Junping Peng, Can-Li Song, Shuaihua Ji, Xi Chen, Yayu Wang, Lin Gu, Xingjiang Zhou, Mingwei Chen, Jian Wang, Dung-Hai Lee, Fuchun Zhang, Xincheng Xie, Shengbai Zhang, and Jin-Feng Jia. Ding Zhang provided assistance in preparing the figures. This work was supported by the National Natural Science Foundation of China and the National Basic Research Program of China.

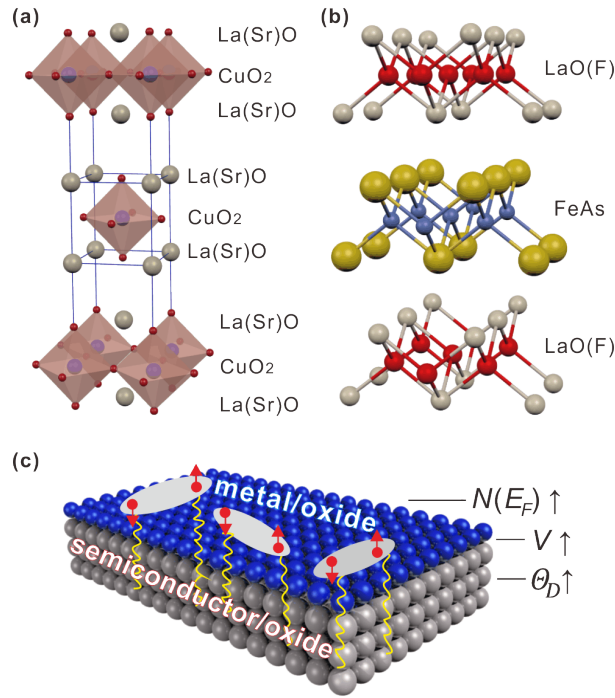

Fig.1 (a) and (b) Schematic layered structure of La$_2$CuO and LaOFeAs, respectively. (c) Schematic of metal/oxide heterostructure. The pairing of electrons (gray ellipse covering two red dots with the arrows showing the spin) in epitaxial films or at interface is mediated by coupling to phonons (yellow curves) from semiconductor/oxide substrate.

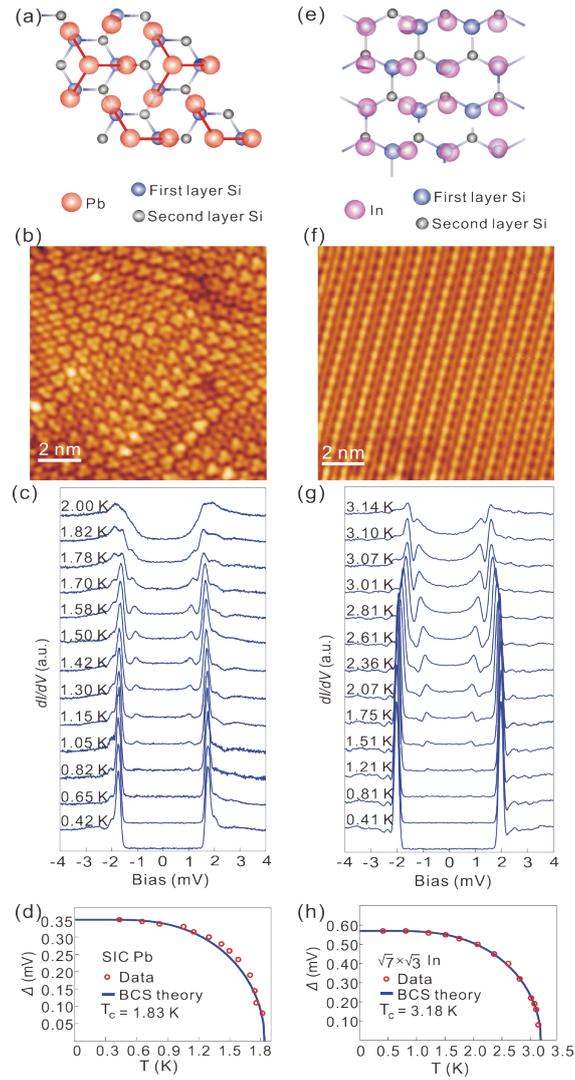

Fig.2 (a) and (e) Schematic structure models, (b) and (f) high resolution STM images ($V = 100$ mV, $I = 50$ pA) of SIC-Pb/Si(111) and $\sqrt{7} \times \sqrt{3} - $In/Si (111), respectively. (c) and (g) the dI/dV spectra measured with a Nb tip as a function of temperature ($V = 10$ mV, $I = 200$ pA), (d) and (h) the superconducting gap as a function of temperature for SIC-Pb/Si(111) and $\sqrt{7} \times \sqrt{3} - $In/Si (111), respectively. Adapted from Zhang *et al.* 2010 *Nat. Phys.* **6**, 104 [29].

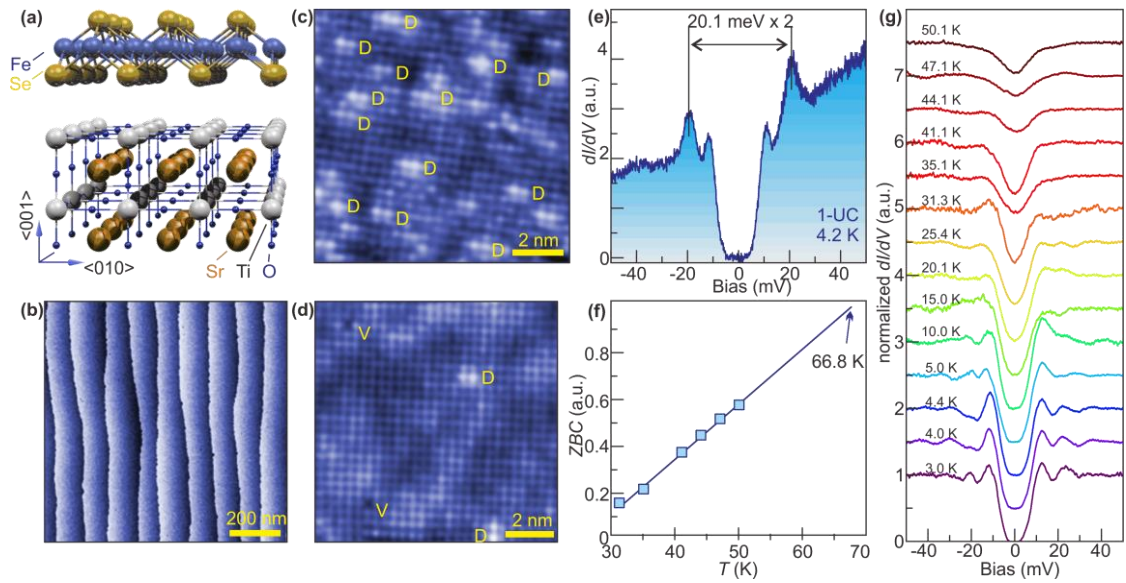

Fig.3 (a) Sketch of 1UC-FeSe/STO(001). (b) STM topography ($V$ = 900 mV, $I$ = 80 pA) of 1UC-FeSe/Nb-STO(001). (c)-(d) Atomically resolved image ($V$ = 100 mV, $I$ = 100 pA) of 1UC FeSe films under Se rich and nearly stoichiometric states, respectively. The D and V in (c) and (d) marks the Se dimers and Se vacancies, respectively. (e) The typical $dI/dV$ spectrum ($V$ = 50 mV, $I$ = 100 pA) taken on the 1UC-FeSe/Nb-STO(001) at 4.2 K revealing the appearance of superconducting gap. (f) The temperature dependence of ZBC extracted from the $dI/dV$ spectra in (g). (g) A series of $dI/dV$ spectra ($V$ = 50 mV, $I$ = 100 pA) taken at various temperatures on 1UC-FeSe/STO(001). (e) adapted from Wang *et al*. 2012 *Chin. Phys. Lett.* **29**, 037402 [31], and (g) and (f) from Zhang *et al*. 2014 *Phys. Rev. B* **89**, 060506 [49].

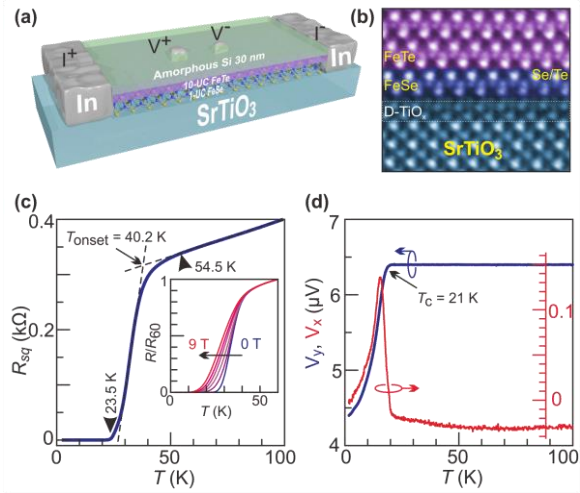

Fig.4 (a) Schematic structure for the transport measurements in the heterostructure of amorphous Si/10UC-FeTe/1UC-FeSe/STO(001). (b) The high-angle annular dark field STEM image of a 10UC-FeTe/1UC-FeSe/STO(001) heterostructure. The dashed rectangle marks the double $TiO_x$ layer. (c) The temperature dependence of resistance, with the inset under various perpendicular magnetic field up to 9 T. (d) The diamagnetic response measured by a homebuilt two-coil mutual inductance system. (b) adapted from Li *et al.* 2016 *2D Materials* **3**, 024002 [48], and (c) and (d) from Zhang *et al.* 2014 *Chin. Phys. Lett.* **31**, 017401 [32].

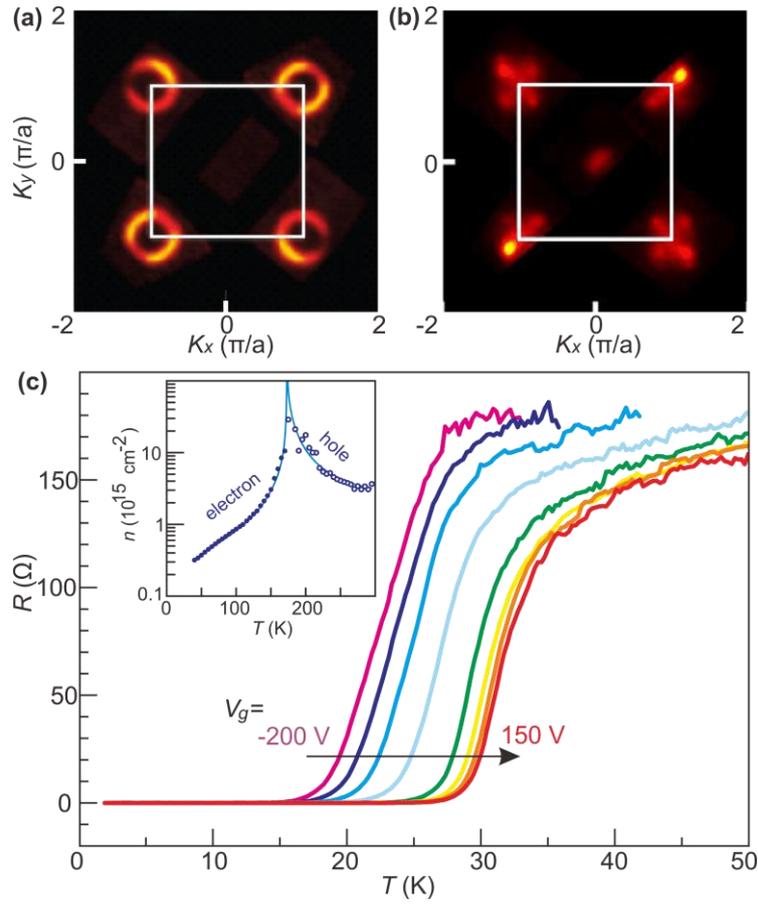

Fig.5 (a)-(b) integrated spectral intensity as a function of momentum for 1UC-FeSe/Nb-STO(001) and 2UC-FeSe/Nb-STO(001), respectively. (c) The *R(T)* curves of 10UC-FeTe/1UC-FeSe/STO(001) at various gate voltages. Inset: carrier density as a function of temperature. (a) adapted from He *et al.*, 2013 *Nat. Mater.* **12**, 605 [52], (b) from Liu *et al.*, 2014 *Nat. Commun.* **5**, 5047 [70], and (c) from Zhang *et al.*, 2014 *Phys. Rev. B* **89**, 060506 [49].

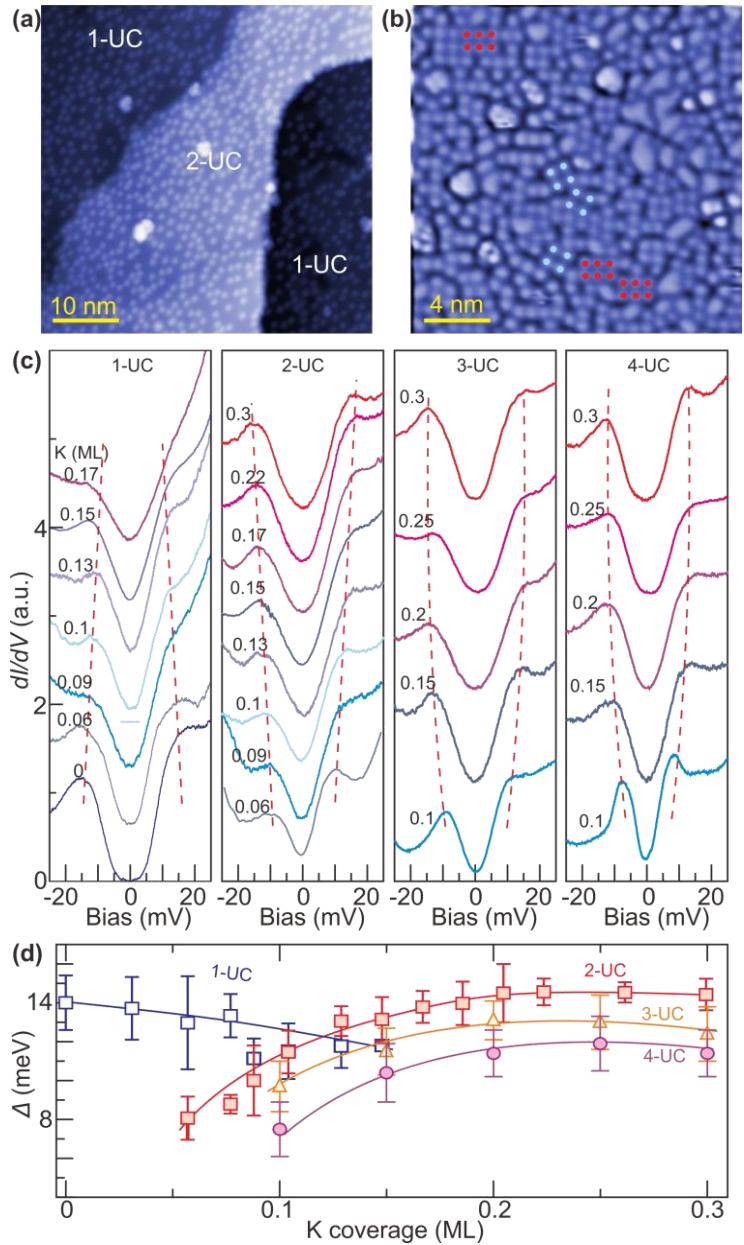

Fig.6 (a) and (b) Topographic images ($V = 1$ V, $I = 50$ pA) of FeSe films on STO(001) after 0.10 ML and 0.16 ML K adsorption, respectively. The red and light blue dots in (b) illustrate the $2 \times 2$ and $\sqrt{5} \times \sqrt{5}$ reconstructions, respectively. (c) The typical $dI/dV$ spectra ($V = 30$ mV, $I = 100$ pA) taken on 1-4 UC FeSe films at various K coverage. The dashes are guide for eyes, showing the change of coherence peaks. (d) The dependence of the superconducting gaps on K coverage. (a) and (b) adapted from Tang *et al.*, 2015 *Phys. Rev. B* **92**, 180507 [75], and (c) and (d) from Tang *et al.*, 2016 *Phys. Rev. B* **93**, 020507 [47].

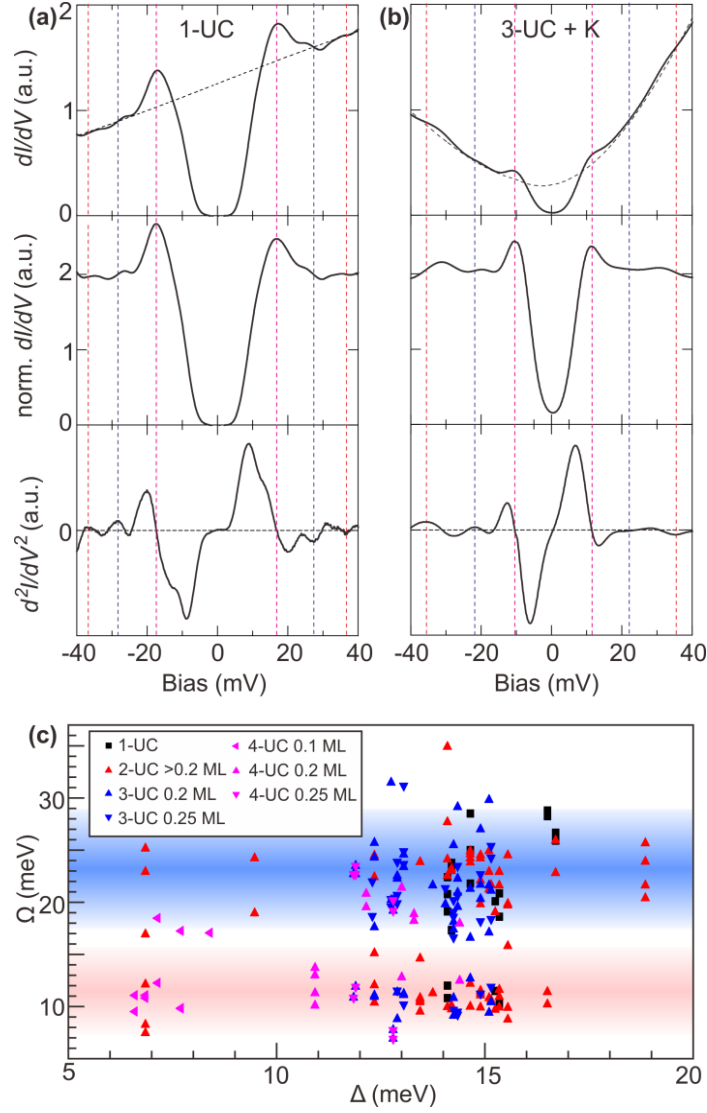

Fig.7 (a) and (b) Black curves show raw *dI/dV* (top panel, *V* = 50 mV, *I* = 100 pA), normalized *dI/dV* (middle panel), and *d²I/dV²* (bottom panel) spectra on bare 1UC-FeSe/Nb-STO(001) and 3UC-FeSe/Nb-STO(001) at the K coverage of 0.20 ML, respectively. The normalization was performed by dividing the raw *dI/dV* spectrum by its background, which was extracted from a cubic fit to the conductance for |*V*| > 20 mV (the dashed line in the top panel). The pink, blue and red dashes show the approximate energy positions of $\pm\Delta$, $\pm(\Delta + \Omega_1)$ and $\pm(\Delta + \Omega_2)$, respectively. (c) The distribution of the phonon energy $\Omega$ as a function of the superconducting gap magnitude $\Delta$. Adapted from Tang *et al.*, 2016 *Phys. Rev. B* **93**, 020507 [47].

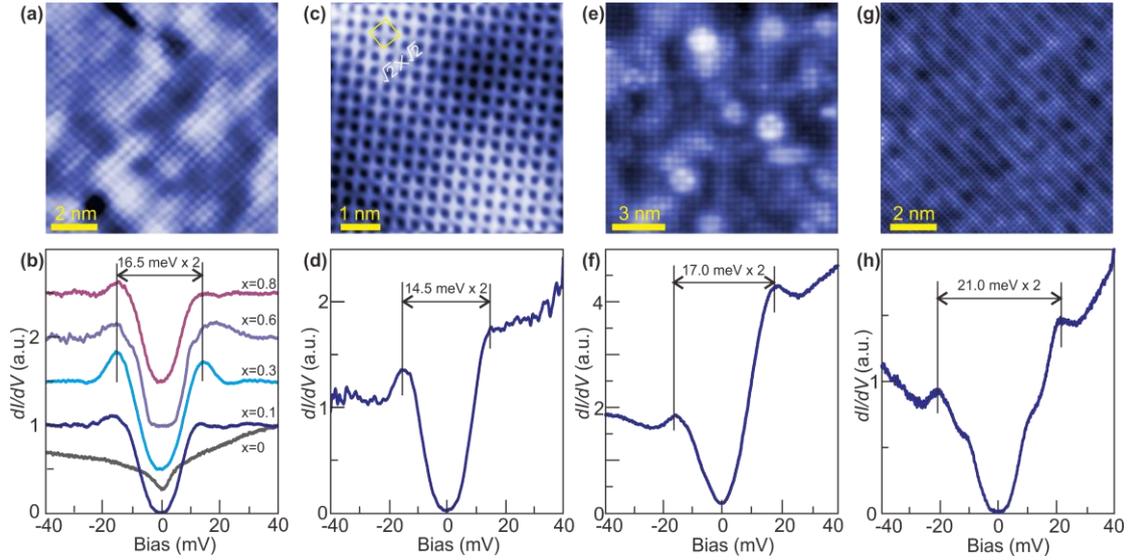

Fig.8 (a), (c), (e) and (g) Atomically resolved images ((a) $V$=200 mV, $I$=100 pA, (c) $V$=100 mV, $I$=50 pA, (e) $V$=70 mV, $I$=300 pA and (g) $V$=50 mV, $I$=100 pA ) of 1UC films of FeTe$_{0.4}$Se$_{0.6}$ on Nb-STO(001), K$_x$Fe$_2$Se$_2$ on Nb-STO(001), FeSe on Nb-STO(110) and FeSe on TiO$_2$(001), respectively. (b), (d), (f) and (h) the corresponding $dI/dV$ spectra ($V$ = 30 mV, $I$ = 100 pA) taken at 4.6 K on the films shown in (a), (c), (e) and (g), respectively. (a) and (b) adapted from Li *et al.*, 2015 *Phys. Rev. B* **91**, 220503 [48], (c) and (d) from Tang *et al.*, 2015 *Phys. Rev. B* **93**, 180507[75], (e) and (f) from Zhou *et al.* 2016 *Appl. Phys. Lett.* **108**, 202603[65], and (g) and (h) from Ding *et al.* 2016 *Phys. Rev. Lett.* **117**, 067001 [64].